\documentclass[a4paper,12pt]{article}

\usepackage{ytableau}\ytableausetup{boxsize=0.7em}
\usepackage{amsmath}
\usepackage{amssymb}
\usepackage{amsfonts}
\usepackage{amsthm}
\usepackage{bbm}
\usepackage{color}
\usepackage{cancel} 
\usepackage{ulem}
\usepackage{cite}
\usepackage{mathalpha} 

\newcommand{\be}{\begin{equation}}
\newcommand{\ee}{\end{equation}}
\newcommand{\bea}{\begin{eqnarray}}
\newcommand{\eea}{\end{eqnarray}}
\newcommand{\bref}[1]{(\ref{#1})}
\newcommand{\nn}{\nonumber}

\newcommand{\mapright}[1]{\smash{[mathop{\hbox to 1cm{\rightarrowfill}}
\limits^{#1}}}

\newcommand{\del}{\partial}

\allowdisplaybreaks[3]
\def\dd{\hbox{\,\Large$\triangleright$}}

 \makeatletter
 \@addtoreset{equation}{section}
 \makeatother

\date{empty}
\begin{document}
\begin{titlepage}
\null
\begin{flushright}
July, 2026 \\
KEK-TH-2814
\end{flushright}
\vskip 2cm
\begin{center}
{\Large \bf 
Mega-Space Current Algebra and
\\
\vspace{0.3cm}
Green-Schwarz Geometry in Heterotic String Theory
}
\vskip 1.5cm
\normalsize
\renewcommand\thefootnote{\alph{footnote}}

{\large
Machiko Hatsuda${}^{*}$\footnote{mhatsuda(at)juntendo.ac.jp}, 
Shin Sasaki${}^{\dagger}$\footnote{shin-s(at)kitasato-u.ac.jp}
and 
Masaya Yata${}^{\ddagger}$\footnote{m-yata(at)juntendo.ac.jp}
}
\vskip 0.5cm
  {\it
  ${}^{*}$
  Department of Radiological Technology, Faculty of Health Science,
  Juntendo University \\
  \vspace{-0.1cm}
  Hongo, Bunkyo-ku, Tokyo 113-0033, Japan \\
  ${}^{*}$KEK Theory Center, High Energy Accelerator Research Organization \\
  \vspace{-0.1cm}
  Tsukuba, Ibaraki 305-0801, Japan \\
  ${}^{\dagger}$
  Department of Physics, Kitasato University, 
  Sagamihara 252-0373, Japan \\
  ${}^{\ddagger}$
  Physics Division, Faculty of Medicine, Juntendo University, Chiba 270-1695, Japan
  }
\vskip 1.0cm
\begin{abstract}
We study the Lorentz and gauge connections in heterotic string theories
within a framework of manifest T-duality. A central motivation is the
Bergshoeff--de Roo (BdR) type parallelism between the torsionful Lorentz
and Yang-Mills connections in heterotic $\alpha'$-corrections.
We formulate worldsheet current algebras in a mega-space that incorporates
the Lorentz and gauge sectors simultaneously. By generalizing the
Pol\'a\v{c}ek-Siegel scheme, we evaluate the commutators of the extended
generators and determine the generalized connections. We show that this
mega-space current algebra geometrically embeds the heterotic
Chern--Simons structures into the generalized vielbeins and curvatures.
We also show that the Green-Schwarz anomaly cancellation condition
follows from the Jacobi identity for the algebra of stringy covariant
derivatives, the ``nachos'' operators $\triangleright_{\mathcal{M}}$.
\end{abstract}

\end{center}

\end{titlepage}

\newpage
\setcounter{footnote}{0}
\renewcommand\thefootnote{\arabic{footnote}}
\pagenumbering{arabic}
\tableofcontents

\section{Introduction} \label{sect:introduction}

A characteristic feature of heterotic string theory is that the Lorentz
and non-Abelian gauge connections enter the $\alpha'$-corrections in
closely parallel forms.
For example, the modified three form field strength appearing in
heterotic supergravity takes the schematic form
\[
\widehat{H}
=
dB
+
\frac{\alpha'}{4}
\big(
\Omega_{\rm YM}
-
\Omega_{\rm L}^{(-)}
\big),
\]
where $\Omega_{\rm YM}$ is the Yang-Mills Chern--Simons form for the
SO(32) or E$_8\times$ E$_8$ gauge connection, while
$\Omega_{\rm L}^{(-)}$ is the Lorentz Chern-Simons form constructed
from the torsionful spin connection $\omega_-$.
This parallel appearance of the gauge and Lorentz sectors is required by
the Green-Schwarz anomaly cancellation mechanism
\cite{Green:1984sg}.
It is also closely related to the Bergshoeff-de Roo (BdR)
identification \cite{Bergshoeff:1988nn,Bergshoeff:1989de}, in which the
torsionful Lorentz connection is treated in close analogy with a
Yang-Mills connection in organizing the gravitational
$\alpha'$-corrections.
At the same time, the two connections have different geometric origins:
the Yang-Mills connection is an internal gauge field, whereas the
Lorentz connection is associated with the tangent bundle of spacetime.
It is therefore desirable to place the Lorentz and gauge sectors in a
common algebraic framework while keeping their geometric origins
explicit.

In a formulation with manifest T-duality, the same question should be
addressed in a duality covariant language.
Since $\alpha'$-corrections to supergravities are constrained by
T-duality \cite{Sen:1991zi}, one should understand how the parallel roles
of the Yang-Mills and torsionful Lorentz connections are encoded in an
O($D$,$D$) covariant or extended geometric framework.
A convenient starting point is an O($D$,$D$) covariant formalism, in which
the ordinary momentum and the winding current are combined into
O($D$,$D$) quantities \cite{Tseytlin:1990nb, Siegel:1993xq, Siegel:1993th,
Siegel:1993bj}.
Double field theory (DFT) \cite{Hull:2009mi} makes this covariance
manifest and has been used to study higher derivative corrections and
generalized Green-Schwarz transformations
\cite{Hohm:2013jaa,Hohm:2014eba,Hohm:2014xsa,Eloy:2020dko,Gitsis:2025clo,Gitsis:2026ufz}.
In this context, the BdR idea has been extended to generalized
Bergshoeff-de Roo identifications in duality covariant formulations
\cite{Baron:2018lve,Baron:2020xel,Gitsis:2024gfb}.
Extended versions of DFT often describe the relevant geometry in an
enlarged space, sometimes referred to as mega-space
\cite{Butter:2022iza}.

Recently, there has been renewed interest in defining the geometry of
DFT and its extensions in such enlarged spaces.
In generalized geometry \cite{Coimbra:2014qaa} and gauged DFT
\cite{Hohm:2011ex,Grana:2012rr}, enlarged spaces provide a useful way
to organize tensor hierarchy structures and gauge group embeddings \cite{Hassler:2025rag}.
Recent works have also shown that all order Green-Schwarz
transformations can be derived systematically in mega-space, giving a
powerful description of the tensor hierarchy
\cite{Hassler:2023axp,Gitsis:2025clo}.
These developments indicate that mega-space provides a natural setting
for describing Green-Schwarz type structures in a duality covariant
language.

In this paper, we take a complementary route based on the
Pol\'a\v{c}ek-Siegel current algebra \cite{Polacek:2013nla}.
This construction gives an O($D$,$D$) covariant current algebraic origin
for the geometry of manifest T-duality \cite{Osten:2019ayq, Hassler:2024hgq}.
In this formulation, the relevant geometry is not introduced by
postulating a generalized connection on the doubled tangent bundle.
Instead, it is obtained from the affine current algebra associated with
the left and right Poincar\'{e} groups.
The basic currents include not only the doubled translation generators,
but also the Lorentz generators and their nondegenerate partners.
This enlargement is essential: the additional currents make the affine
algebra consistent and allow the Lorentz connection to appear as a
component of the generalized vielbein.
Thus the spin connection is not added by hand to a covariant derivative,
but is already part of the stringy background seen by the worldsheet
currents.

Our aim is to extend this framework by adding the heterotic
non-Abelian gauge currents to the same enlarged algebra.
We show that, after imposing the torsion constraints and the physical
section condition, the resulting mega-space geometry reproduces the
Yang-Mills and Lorentz structures that enter the heterotic
Green-Schwarz mechanism.
In this way, the Lorentz and gauge connections are placed in a common
current algebraic framework, while their different geometric origins
remain explicit.

The organization of this paper is as follows.
In Section \ref{sec:Lorentz_connections}, we briefly review the Lorentz connection and curvature in
the extended space following Pol\'a\v{c}ek and Siegel \cite{Polacek:2013nla}.
In Section \ref{sect:NddP}, we extend this formalism to include the heterotic
non-Abelian gauge sector.
In Section \ref{sec:geometry}, we examine the current algebra associated with the
extended vielbeins, impose the torsion constraints in mega-space, and
derive the corresponding effective gravity theory.
We also show that the Green-Schwarz condition follows from the Bianchi
identity.
Section \ref{sec:conclusion} is devoted to conclusions and discussions.

\section{Lorentz connections in doubled formalism} \label{sec:Lorentz_connections}

In this section we introduce the Pol\'{a}\v{c}ek and Siegel scheme of the
extended connection in a doubled space \cite{Polacek:2013nla}.
We first focus on  the Lorentz sector in an O($D$,$D$) covariant doubled
formalism 
{after the one in the $D$-dimensional spacetime.}
The gauge sector in heterotic theories will be introduced in the next section.
The basic idea is to formulate the geometry of manifest T-duality in
terms of an affine current algebra
{where the Lorentz connection is a part of the background vielbein field in the extended space.
As in ordinary gauge theories, the curvature tensor is obtained from the commutator of covariant derivatives and can be interpreted as the corresponding field strength.
The symmetry transformation rule is derived from the commutator of the gauge transformations generated by parameter vectors in the extended space.}

\par
\vspace{6 mm}
{
\subsection{Lorentz connection in the $D$-dimensional spacetime}}
\label{subsection:21}

{
General relativity has been formulated in a Lorentz covariant manner (for example section IX of \cite{Siegel:1999ew}), thereby clarifying the role of the connection as a Lorentz connection.
The Lorentz connection $\omega_a{}^{mn}$ defines the covariant derivative
\bea
\nabla_a&=&e_a{}^m p_m+\frac{1}{2}\omega_a{}^{mn}S_{mn} \label{covderiLorp}
\eea
as the gauge field for the Lorentz symmetry generated by 
\bea
&\left[S_{mn},S_{lk}\right]~=~\frac{1}{i}
(\eta_{km} S_{nl}-\eta_{kn} S_{ml}-\eta_{lm} S_{nk}+\eta_{ln} S_{mk})
~,~
\left[S_{mn},p_{l}\right]~=~\frac{1}{i}
(\eta_{ln} p_{m}-\eta_{lm} p_{n})~~.&
\eea
A gauge $S_{ab}=\frac{1}{2}\delta_a^{[m}\delta_b^{n]} S_{mn}$ is chosen
for simpler computation in
the local Lorentz indices $a,b,c,\cdots$,
then $\frac{1}{2}\omega_a{}^{bc}S_{bc}$ is used instead of $\frac{1}{2}\omega_a{}^{mn}S_{mn}$.  
Riemannian curvature tensor $R_{abcd}$ is obtained from the 
commutator of the covariant derivatives \bref{covderiLorp}
\bea
\left[\nabla_a,\nabla_b\right]&=&\frac{1}{i}\left(
T_{ab}{}^c\nabla_c+\frac{1}{2}R_{ab}{}^{cd} S_{cd}\right)\nn\\
&&\left\{ 
\begin{array}{ccl}
	T_{ab}{}^c &=& c_{ab}{}^c-\omega_{[ab]}{}^c~~,~~c_{ab}{}^c=
	e_{[a|}{}^m(\partial_me_{|b]}{}^n)e_n{}^c \\
	R_{ab}{}^{cd} &=& e_{[a|}{}^{m}\partial_{m}\omega_{|b]}{}^{cd}
	+\omega_{[a|}{}^{ce}\omega_{|b]}{}^d{}_e
	-\omega_{[ab]}{}^e\omega_e{}^{cd} 
\end{array}\right.\label{RLr}
\eea
Torsion free condition relates the Lorentz connection $\omega_{abc}$ to the vielbein field $e_a{}^m$ as
\bea
&&T_{ab}{}^c~=~0~~\to~
\left\{
\begin{array}{ccl}
	c_{abc}+\omega_{bac}-\omega_{abc}&=&0\cdots (1)\\
	c_{bca}+\omega_{cba}-\omega_{bca}&=&0\cdots (2)\\
	c_{cab}+\omega_{acb}-\omega_{cab}&=&0\cdots (3)\\
\end{array}
\right.
\nn\\
&&(1)-(2)-(3)=0\nn\\
&&\Rightarrow~~\omega_{abc}~=~
\frac{1}{2}(c_{abc}+c_{c(ab)})~=
\frac{1}{2}(c_{a[bc]}-c_{bca})
~~.\label{Aomega}
\eea
The trace of the Lorentz connection $\omega_{abc}$ is given by
\bea
\omega^b{}_{ab}&=&\frac{1}{\bf e}\partial_m(e_a{}^m {\bf e})~~,~~{\bf e}={\rm det}~ e_m{}^a~~~. 
\eea
}

{
	The rank four tensor, which has the symmetry of indices 
	$R_{abcd}=-R_{bacd}$ and $R_{abcd}=-R_{abdc}$ from \bref{RLr},
	is decomposed as
	\bea
	&\raisebox{0.7ex}
	{\ydiagram{1,1}}~~\otimes~~ 
	\raisebox{0.7ex}
	{\ydiagram{1,1}}
	~~=~~
	\raisebox{2.5ex}
	{\ydiagram{1,1,1,1}}
	~~\oplus~~
	\raisebox{1.5ex}
	{\ydiagram{2,1,1}}
	~~\oplus~~
	\raisebox{0.7ex}
	{\ydiagram{2,2}}
	&\nn
	\eea
	The tensor product of rank two tensors with the dimension $\frac{1}{2}D(D-1)$ is sum of the totally antisymmetric part
	$\frac{1}{4!}D(D-1)(D-2)(D-3)$, $\frac{1}{4\cdot 2}(D+1)D(D-1)(D-2)$,
	and the mixed symmetry $
	\frac{1}{3\cdot 2^2}(D+1)D^2(D-1)$.
	The Bianchi identity, $\left[\nabla_{[a},\left[\nabla_b,\nabla_{c]}\right]\right]=0$, implies  $R_{[abc]d}=0$ and $D_{[a}R_{bc]de}=0$ with $[i\nabla_a,\Phi]=D_a\Phi$.
The condition $R_{[abc]d}=0$ together with the antisymmetry property of the indices $ab$ and $cd$ of $R_{abcd}$
determines
	the physical Riemannian tensor which corresponds to  $\raisebox{0.7ex}
	{\ydiagram{2,2}}$, $R_{abcd}=R_{cdab}$.}

{The symmetry transformation rules are derived from the commutator
	of the parameter vector.
	The Lie derivative and the local Lorentz transformation are generated by the parameter vector in the Poincar\'{e} space $\Xi^M\nabla_M$ with $\Xi^M=(\xi^m,\lambda^{mn})$. Again  $\frac{1}{2}\lambda^{ab}S_{ab}$ is used instead of $\frac{1}{2}\lambda^{mn}S_{mn}$ for simpler computation in $\Xi^M\nabla_M=\xi^m p_m+\frac{1}{2}\lambda^{ab}S_{ab}$.
	\bea
	&&\left[\Xi^M\nabla_M,\nabla_a\right]~=~
	\frac{1}{i}\left(\delta e_a{}^m p_m+\frac{1}{2}\delta \omega_{a}{}^{bc}S_{bc}\right)\nn\\
	&&~~~~~~~~~ \left\{
	\begin{array}{ccl}
		\delta e_a{}^m &=&\xi^n\partial_n e_a{}^m -e_a{}^n\partial_n\xi^m\\
		\delta \omega_a{}^{bc}&=&\xi^m\partial_m \omega_a{}^{bc}
		-e_a{}^m\partial_m \lambda^{bc}+\omega_{d}{}^{bc}\lambda^d{}_a
		+\omega_{ad}{}^{c}\lambda^{db}
		+\omega_{a}{}^{b}{}_d\lambda^{dc}
	\end{array}
	\right. \label{gaugetransfL}
	\eea
	The Lorentz connection $\omega_{a}{}^{bc}$ is the gauge field associated with local Lorentz symmetry, as is evident from the second term in the gauge transformation law \bref{gaugetransfL}.
}

{
	In the standard  general relativity the Christoffel symbol $\Gamma^{m}{}_{nl}$ defines the covariant derivative
	\bea
	\nabla_m&=&p_m-\Gamma^n{}_{ml}G_n{}^l
	\eea 
	as the gauge field for the general linear symmetry generated by
	\bea
	&\left[G_n{}^l,G_m{}^k\right]~=~\frac{1}{i}(\delta_m^l G_n{}^k
	-\delta_n^k G_{m}{}^l)~,~
	\left[G_n{}^l,p_m\right]~=~\frac{1}{i}\delta_m^l p_n ~~.&
	\eea
	The Riemann curvature tensor $R_{mn}{}^l{}_k$ is obtained from the commutator of covariant derivatives, as follows.
	\bea
	\left[\nabla_m,\nabla_n\right]&=&\frac{1}{i}\left(
	T_{mn}{}^l\nabla_l+R_{mn}{}^l{}_k G_l{}^k\right)\nn\\
	&&\left\{ 
	\begin{array}{ccl}
		T_{mn}{}^l&=&\Gamma^l{}_{[mn]}\\
		R_{mn}{}^l{}_k&=&\partial_{[m|}\Gamma^l{}_{n]k} 
		+\Gamma^l{}_{[m|r}\Gamma^r{}_{|n]k}
	\end{array}
	\right.\label{RGL}
	\eea
	Torsion free condition is given by  
	\bea
	&T_{mn}{}^l~=~0~~\Rightarrow~~\Gamma^l{}_{[mn]}~=~0~~~.&\label{Chris+}
	\eea
	The metric compatible condition with   \bref{Chris+} gives
	\bea
	&\nabla_mg_{nl}~=~0~~\Rightarrow~~\Gamma^m{}_{nl}~=~
	\frac{1}{2}g^{mr}(-\partial_rg_{nl}
	+\partial_ng_{rl}+\partial_lg_{rn})~~~.
	&\label{Chris}
	\eea
	The trace of the Christoffel symbol $\Gamma^m{}_{nl}$ is given as
	\bea
	\Gamma^n{}_{mn}&=&\partial_m {\rm log} \sqrt{-g}~~,~~g={\rm det}~ g_{mn}~~~.
	\eea 
}

{
The relation between the Lorentz connection
$\omega_{ab}{}^c$ and the  Christoffel symbol $\Gamma^{l}{}_{mn}$ is determined as follows for a vector $V_a=e_a{}^mV_m$,
\bea
&&\nabla_m V_{n}~=~e_m{}^ae_n{}^b(\nabla_aV_b)~~
\nn\\
&&\displaystyle
\left\{
\begin{array}{ccl}
\nabla_m V_{n}&=&\left(\displaystyle\frac{1}{i}\partial_m-\Gamma^r{}_{ml}G_r{}^l\right)V_n
=\displaystyle\frac{1}{i}(\partial_mV_n-\Gamma^r{}_{mn}V_r)\\
\nabla_aV_{b}&=&
\left(\displaystyle\frac{1}{i}e_a{}^m\partial_m+\frac{1}{2}\omega_a{}^{cd}S_{cd}\right) V_b
=\displaystyle\frac{1}{i}(e_a{}^m\partial_mV_b+\omega_a{}^c{}_b V_c)
\end{array}
\right.\nn\\
&&\Rightarrow
~~\Gamma^k{}_{mn}=e_m{}^ae_n{}^b\omega_{ab}{}^c e_c{}^k+
(\partial_m e_n{}^a)e_a{}^k~~~.
\label{Chrisomega}
\eea
The Riemannian tensor in \bref{RLr} is related to the Riemann tensor \bref{RGL} as
\bea
R_{abcd}&=&-e_a{}^m e_b{}^n e_{lc}e_d{}^kR_{mn}{}^l{}_k~~~.
\eea
}

\par
\vspace{6 mm}
\subsection{Lorentz connection in the doubled formulation}
\label{subsection:22}

In the doubled formulation the
momentum and the winding currents are combined into an
O($D$,$D$) vector  $P_M (\tau,\sigma)= (P_m,~P_{m'})$ in the left/right index, while the Lorentz 
generators $S_{MN}(\tau, \sigma)=(S_{mn},~S_{m'n'})$ are kept as part of the
affine current algebra.
{Here $(\tau,\sigma)$ are the worldsheet coordinates.
In the following, we omit the $\tau$ dependence in the algebras at equal times.}
The concrete representation of the currents reduces to the  following
expression in terms of the string coordinates in $D$ dimensions
$X^{\underline{m}}(\tau,\sigma)$ as :
$\frac{1}{\sqrt{2}}(P_{m}+P_{m'})=\frac{1}{i}\partial_{\underline{m}}$ and
$\frac{1}{\sqrt{2}}(P_{m}-P_{m'})=\partial_\sigma X^{\underline{m}}$.

\par
\vskip 6mm 
\subsubsection{``Dual'' Lorentz generator $\Sigma^{MN}$}
The momentum $P_{{M}}$ with the double space index $M$ satisfies the
algebra\footnote{
We employ the convention $2 \pi \alpha^{\prime} = 1$ where $\alpha'$
is the string Regge parameter which will be specified when necessary.
} 
with the O$(D,D)$ invariant metric	$\eta_{MN}$
\begin{align}
[P_{{M}} (\sigma), P_{{N}}
 (\sigma')] = 
i \eta_{{M}{N}} \del_{\sigma} \delta (\sigma - \sigma'),\label{PPSch}
\end{align}
A naive extension of the Poincar\'{e} algebra by the inclusion of this Schwinger term violates the Jacobi identity of $(P,P,S)$.
Let  ${\cal J}$  be the totally antisymmetric sum of the commutators of  ($P_M(\sigma),~P_N(\sigma'), ~S_{LK}(\sigma'')$)
	\bea
	{\cal J} 
	&=&\left[\left[   P_M(\sigma),P_N(\sigma') \right],S_{LK}(\sigma'') \right] +
	\left[\left[  P_N(\sigma'), S_{LK}(\sigma'') \right] ,  P_M(\sigma)\right] \nn\\&&+
	\left[\left[  S_{LK}(\sigma''), P_M(\sigma)\right] ,P_N(\sigma')\right] ~~.\label{Jacobilhs}
	\eea
This Jacobi identity is not satisfied due to the Schwinger term in \bref{PPSch}, as
	\bea
	&&\Rightarrow{\cal J}~=~\eta_{M[L}\eta_{K]N}\big(\delta(\sigma'-\sigma'') \partial_\sigma\delta(\sigma-\sigma')-
	\delta(\sigma-\sigma'') \partial_\sigma\delta(\sigma-\sigma')\big)\neq 0~~~\nn\\
	&&~~~\because~  \displaystyle\int \int \int d\sigma d\sigma' d\sigma'' f(\sigma)g(\sigma')h(\sigma'')
	{\cal J}=-\eta_{M[L}\eta_{K]N}\int d\sigma \partial_\sigma(fg)h~\neq~0~~~.
	\eea
In the last line above the hyperfunction is evaluated as the integral with arbitrary test functions $f(\sigma)$, $g(\sigma')$, $h(\sigma'')$.

	To cancel out the Schwinger term from $(P,P)$ in \bref{PPSch}, 
	the dual Lorentz generator $\Sigma^{MN}(\sigma)$ is introduced so that its commutator with the Lorentz generator $S_{MN}(\sigma)$ produces an additional Schwinger term in $(S,\Sigma)$.
The non-degenerate Poincar\'{e} algebra  is written as
\bea
&&{\renewcommand{\arraystretch}{1.5}
	\left\{\begin{array}{ccl}
	\left[S_{MN}(\sigma),S_{LK}(\sigma')\right]&=&-i\eta_{[K|[M}S_{N]|L]}\delta(\sigma-\sigma')\\
		\left[S_{MN}(\sigma),P_{L}(\sigma')\right]&=&-iP_{[M}\eta_{N]L}\delta(\sigma-\sigma')\\
		\left[P_{M}(\sigma),P_{N}(\sigma')\right]&=&i
\eta_{ML}\eta_{NK}\Sigma^{LK}\delta(\sigma-\sigma')+ i\eta_{MN}\partial_\sigma\delta(\sigma-\sigma')\\
		\left[S_{MN}(\sigma),\Sigma^{LK}(\sigma')\right]&=&-i\delta^{[K}_{M}\Sigma_{N}^{~~L]}\delta(\sigma-\sigma')-i\delta^{[K}_M\delta_N^{L]}\partial_\sigma\delta(\sigma-\sigma') 
	\end{array}
	\right.}~~ \label{ndPMN}
\eea
and the others are zero.
$\Sigma^{LK}(\sigma)$  ensures that
	the Jacobi identity holds, ${\cal J}=0$, as
	\bea
	{\cal J}&=&\eta_{M[L}\eta_{K]N}\nn\\
	&&\times\big(\delta(\sigma-\sigma') \partial_{\sigma''}\delta(\sigma''-\sigma)-
	\delta(\sigma-\sigma'') \partial_\sigma\delta(\sigma-\sigma')
	-\delta(\sigma'-\sigma'') \partial_{\sigma'}\delta(\sigma'-\sigma)
	\big)=0~~~\nn\\
	&\because& \displaystyle\int \int \int d\sigma d\sigma' d\sigma'' f(\sigma)g(\sigma')h(\sigma''){\cal J} =-
	\eta_{M[L}\eta_{K]N}
	\int d\sigma \partial_\sigma(fgh)~=~0~~~.
	\eea
The non-degenerate dual partner 
$\Sigma^{MN}$ was firstly introduced in \cite{Hatsuda:2001xf}
as the Lorentz current of the AdS superstring.

The affine non-degenerate Poincar\'{e} algebra in \bref{ndPMN} 
is generated by $2D^2$-dimensional basis,  
sometimes called ``nachos'' operators:
\begin{align}
\dd_{\mathcal{M}}
= 
\Big(
S_{{MN}}, 
P_{{M}}, 
\Sigma^{{MN}}
\Big).
\end{align}
Using this basis, the algebra is recast into the form,
\begin{align}
\Big[
\dd_{\mathcal{M}}(\sigma), 
\dd_{\mathcal{N}} (\sigma')
\Big]
= i \eta_{\mathcal{M} \mathcal{N}} \del_{\sigma} \delta (\sigma -
 \sigma') 
- i f_{\mathcal{M} \mathcal{N}} {}^{\mathcal{P}}
\dd_{\mathcal{P}} \, \delta (\sigma - \sigma')~~~.
\label{eq:flat_current_algebra}
\end{align}
The invariant metric of the non-degenerate doubled Poincar\'{e} group 
$\eta_{\mathcal{MN}}$ is defined by
\bea
&&\begin{array}{l}
~~~~~~~~~~~~~~~~~~~_{S}~~~~~~_{P}~~~~~~_{\Sigma}\\
\eta_{\cal MN}=\begin{array}{c}
	_S\\_P\\_{\Sigma}\end{array}
\left(\begin{array}{ccc}
&&\eta_{S\Sigma}\\
&\eta_{PP}&\\
\eta_{\Sigma S}&&\end{array}
\right)=
\left(\begin{array}{ccc}
	&&{\bf 1}\\
	&{\bf 1}&\\
	{\bf 1}&&\end{array}
\right)
~~~.\label{etaSPSig}
\end{array}
\eea
The $f_{\cal MN}{}^{\cal L}$ is the structure constant of the non-degenerate doubled Poincar\'{e} algebra with non-Abelian gauge symmetry. 
In order to ensure the closure of the Jacobi identity of the current algebra
the structure constant must be totally anti-symmetric,
\bea
f_{\cal MNL}\equiv f_{\cal MN}{}^{\cal K}\eta_{\cal KL}=\frac{1}{3!}f_{\cal [MNL]}~~~\label{fabc}
\eea
with ${\cal O}_{[{\cal MN}]}={\cal O}_{\cal MN}- {\cal O}_{\cal NM}$.
It was explained in  \cite{Hatsuda:2014qqa} for the super-Poincar\'{e} group.
The non-zero components are  $f_{SPP}$ and
$f_{SS\Sigma}=f_{SS}{}^S$.

\par
\vskip 6mm 
\subsubsection{Virasoro algebra}
The Virasoro constraint is the O($D$,$D$) invariant current bilinear form as
\bea
{\cal H}_\sigma&=&\displaystyle \frac{1}{2} \dd_{\cal M}\eta^{\cal{MN}}\dd_{\cal N}
=\displaystyle \frac{1}{2}P_{\underline{M}}\eta^{\underline{MN}}P_{\underline{ N}}+
\displaystyle \frac{1}{2}
\Sigma^{MN}S_{MN}~~~.\label{Virsig}
\eea
The Virasoro algebra of the $\sigma$ component Virasoro constraint \bref{Virsig} is given by
\bea
\left[{\cal H}_\sigma(\sigma),{\cal H}_\sigma(\sigma')\right]&=&
i({\cal H}_\sigma(\sigma)+{\cal H}_\sigma(\sigma'))\partial_\sigma\delta(\sigma-\sigma')
~~~.
\label{Vira}
\eea

The worldvolume derivative is generated by the Virasoro constraint ${\cal H}_\sigma$ as
\bea
\partial_\sigma \Phi(X(\sigma))&=&\left[
i\displaystyle \int {\cal H}_\sigma, \Phi(X(\sigma)) \right]
~=~\left(D_{\underline{M}}\Phi(X)\right)~\eta^{\underline{MN}}P_{\underline{N}}+\frac{1}{2}
(D_{MN}\Phi)\Sigma^{MN}~~~,\label{dsigma}
\eea
while the enlarged spacetime derivative is generated by  $\dd_{\cal M}$ as
\bea
D_{\cal M}\Phi(X) &=&i\left[\dd_{\cal M}, \Phi(X)\right]~~~
\eea
for an arbitrary tensor function $\Phi(X(\sigma))$.
This worldvolume derivative is a generalization of the ordinary string in Hamiltonian formulation:
For the string coordinate $x^m$ and the momentum $p_m$ the Virasoro constraint is ${\cal H}_\sigma=\partial_\sigma x^m p_m$
which gives $\partial_\sigma \Phi(x)=\left[i{\cal H}_\sigma,\Phi\right]=\partial_\sigma x^m\partial_m\Phi$.
In this paper we consider only the operators $D_S$ and $D_P$.

The section condition is the zero mode part of the Virasoro constraint which is background independent. 
It acts  on the spacetime functions $\Phi(X)$ and $\Psi(X)$ as
\bea
\partial_{\cal M}\eta^{\cal MN}\partial_{\cal N}\Phi(X)~=~0~=~
\left(\partial_{\cal M}\Psi(X)\right)\eta^{\cal MN}\left(\partial_{\cal N}\Phi(X)\right)~~~,
\eea 
where $\partial_{\cal M} $ is the  derivative with respect to the enlarged space coordinates including the ones  for $S$ and  $\Sigma$ \cite{Hatsuda:2015cia}
as well as the non-Abelian coordinate $y^\mu$ in \cite{Hatsuda:2022zpi}.
\par
\vskip 6mm
\subsubsection{Background fields}

We next introduce gravitational background fields.
The basis 
$\dd_{\mathcal{M}}$ is replaced with 
$
\dd_{\mathcal{A}}
= E_{\mathcal{A}} {}^{\mathcal{M}} 
\dd_{\mathcal{M}}
$ 
where the generalized vielbein $E_{\mathcal{A}}
{}^{\mathcal{M}}$ is given by
\begin{align}
	E_{\mathcal{A}} {}^{\mathcal{M}} 
	=
	\left[
	\begin{array}{ccc}
		\delta_{AB} {}^{MN} &0
		& 0 \\
		\omega_{A} {}^{MN} &
		E_{{A}}{}^{{M}} & 0 \\
		r^{ABMN} -
		\frac{1}{2} \omega^{CAB}
		\omega_{CMN} &
		-\omega^{MAB} & \delta^{AB}
		{}_{MN}
	\end{array}
	\right].\label{Emega}
\end{align}
Here $\mathcal{A}$ and ${A}, {B}, {C},
\ldots$ are the mega-space indices including the local Lorentz sector 
while $\mathcal{M}$ and ${M},
{N}, {L}, \ldots$ are 
the indices including the GL($D$) tensor sector.
The unit operator for the tensor index
$\delta_{AB} {}^{MN} = 
\delta_{[{A}} {}^{{M}} \delta_{{B}]} {}^{{N}}$, the doubled vielbein $E_A{}^M$ and the Lorentz connection $\omega_{A}{}^{MN}$ are introduced.
The orthogonality allows the ambiguity of the antisymmetric tensor
$
r^{ABCD} 	+r^{CDAB} = 0$.

The generalized vielbein satisfies the orthogonality condition:
\begin{align}
E_{\mathcal{A}} {}^{\mathcal{M}} \eta_{\mathcal{MN}}
 E_{\mathcal{C}} {}^{\mathcal{N}} = \eta_{\mathcal{AC}}.
\end{align}
Then in the curved space, the current algebra \eqref{eq:flat_current_algebra} becomes 
\begin{align}
\Big[
\dd_{\mathcal{A}} (\sigma), 
\dd_{\mathcal{B}} (\sigma')
\Big] = i {\eta_{\cal AB}}
  \del_{\sigma}
 \delta (\sigma - \sigma')
- i T_{\mathcal{A} \mathcal{B}} {}^{\mathcal{C}}
\dd_{\mathcal{C}} \delta (\sigma - \sigma').
\end{align}
Here $T_{\mathcal{A} \mathcal{B}} {}^{\mathcal{C}}$ is the generalized
torsion given by
\begin{align}
T_{\mathcal{A} \mathcal{C}} {}^{\mathcal{E}} =
E_{[\mathcal{A}} {}^{\mathcal{M}} 
(
D_{\mathcal{M}} E_{\mathcal{C}]} {}^\mathcal{N}
)
E^{-1}{}_{\mathcal{N}} {}^{\mathcal{E}}
+
\frac{1}{2}
\eta^{\mathcal{E} \mathcal{D}} E_{\mathcal{D}} {}^{\mathcal{M}}
(
D_{\mathcal{M}} E_{[\mathcal{A}|} {}^{\mathcal{N}}
)
E^{-1}{}_{\mathcal{N}} {}^{\mathcal{F}} \eta_{\mathcal{F} | \mathcal{C}]}
+
E_{\mathcal{A}} {}^{\mathcal{M}} 
E_{\mathcal{C}} {}^{\mathcal{N}}
E^{-1}{}_{\mathcal{P}} {}^{\mathcal{E}}
f_{\mathcal{M} \mathcal{N}} {}^{\mathcal{P}}.
\end{align}

{
The gauge transformation rule is obtained by the commutator with the 
parameter vector in the mega-space $\Xi^{\cal M}\dd_{\cal M}$ 
with $\Xi^{\cal M}=(\Xi^S,\Xi^P,\Xi^\Sigma)=(\lambda^{MN},\xi^M,0)$ 
analogously to \bref{gaugetransfL}. 
\bea
\left[\Xi^{\cal M}\dd_{\cal M}(\sigma),\dd_{\cal A}(\sigma')\right]
&=&\frac{1}{i}\delta E_{\cal A}{}^{\cal M}
\dd_{\cal M}\delta(\sigma-\sigma')
+{i}\Xi^{\cal M}E_{\cal AM}(\sigma')
\partial_\sigma\delta(\sigma-\sigma')\nn\\
\delta E_{\cal A}{}^{\cal M}&=&
\Xi^{\cal N}D_{\cal N}E_{\cal A}{}^{\cal M}-
E_{\cal A}{}^{\cal N}(D_{\cal N}\Xi^{\cal M}-
D^{\cal M}\Xi_{\cal N})
\eea
where indices are raised and lowered with $\eta_{\cal  MN}$ or $\eta^{\cal MN}$ and $\left[i\dd_{\cal M},\Phi\right]=D_{\cal M}\Phi$.
The unphysical $\Sigma$-derivative is set to be zero, $D_{\Sigma}=0$,
and the doubled space derivative is denoted as  $D_P=\partial_M$.
The gauge transformation rules in components \bref{Emega}
 are as follows
\bea
\left\{  \begin{array}{ccl}
\delta E_{A}{}^{M}&=&
\xi^{N}\partial_{N}E_{A}{}^{M}-
E_{A}{}^{N}(\partial_{N}\xi^{M}- \partial^{M}\xi_{N})+E_B{}^M\lambda^B{}_A
\nn\\
\delta \omega_{A}{}^{BC}&=&
\xi^{N}\partial_{N}\omega_{A}{}^{BC}-
E_{A}{}^{N} \partial_{N}\lambda^{BC}
+\omega_{D}{}^{BC}\lambda^D{}_A
+\omega_{A}{}^{D[B}\lambda^{C]}{}_D~~~.
\end{array}\right.
\eea
It is useful to note that
\bea
\delta \omega_{M}{}^{BC}&=&
\xi^{N}\partial_{N}\omega_{M}{}^{BC}+
( \partial_{M}\xi^{N}-\partial^N\xi_M)\omega_{N}{}^{BC}
-\partial_M\lambda^{BC}
+\omega_{M}{}^{D[B}\lambda^{C]}{}_D~~~.
\eea
}

\section{Heterotic mega-space current algebra} \label{sect:NddP}

The heterotic string with manifest T-duality is described by 
the set of currents; the momentum, the winding mode and the chiral current 
\cite{Siegel:1993th, Siegel:1993bj}.
This current algebra has recently been used to construct integrable
deformations of heterotic sigma models \cite{Osten:2023cza}.
In this section, we extend the Pol\'{a}\v{c}ek-Siegel
 formalism in the previous section to include non-Abelian gauge sectors.
Following the gauged DFT scheme \cite{Grana:2012rr}, 
we now extend the O($D$,$D$) momentum to the one for O($D+n$,$D$):
{ where $n=$dim $G$  for a gauge group $G$}
\footnote{
For conventional heterotic string theories, the number of additional
left-moving internal directions is $n=16$. In the following, however, we
keep $n$ arbitrary for generality. 
This allows us to introduce a general gauge group, including the
conventional $\mathrm{E}_8 \times \mathrm{E}_8$ and $\mathrm{SO}(32)$ cases with $\dim G=496$.
}.

The set of currents for the nondegenerate Poincar\'{e} algebra with non-Abelian gauge symmetry is given by 
\bea
\dd_{\cal M}&=&(S_{MN},~P_M,~\Sigma^{MN})\nn\\
&=&(S_{mn},~P_m,~\Sigma^{mn}, ~P_\mu; ~S_{m'n'},~P_{m'},~\Sigma^{m'n'})
\eea
where the second line shows the left and right sectors. 
They satisfy the following current algebra
\bea
\left[\dd_{\cal M}(\sigma),\dd_{\cal N}(\sigma')\right]&=&-if_{\cal {MN}}{}^{\cal L} \dd_{\cal L}\delta(\sigma-\sigma')+ i\eta_{\cal {MN}}\partial_\sigma\delta(\sigma-\sigma')~~~.
\label{CAflat}
\eea
The $\eta_{{\cal MN}}$ is given by \bref{etaSPSig}
and 
$\eta_{MN}$ is the O($D+n$,$D$) invariant metric
\bea
&&\begin{array}{l}
	~~~~~~~~~~~~~~~~~~~~~~~~~~~~_{n}~~~~~~_{\nu}~~~~~~~_{n'}\\
\eta_{PP}=	\eta_{MN}=\begin{array}{c}_m\\_\mu\\_{m'}\end{array}
	\left(\begin{array}{ccc}
		\eta_{mn}&&\\
		&\eta_{\mu\nu}&\\
		&&\eta_{m'n'}\end{array}
	\right)=
	\left(\begin{array}{ccc}
		{\bf 1}&&\\
		&{\bf 1}&\\
		&&-{\bf 1}\end{array}
	\right)~~~.
\end{array}
\eea
Non-zero structure constants are
\bea
f_{SS\Sigma},~f_{SPP},~f_{\mu\nu\rho}~~~.
\eea 
The nondegenerate Poincar\'{e} algebra \bref{ndPMN} is extended to include the non-Abelian generator $P_\mu$ as
\bea
&&{\renewcommand{\arraystretch}{1.3}
\left\{\begin{array}{ccl}
	\left[S_{MN}(\sigma),S_{LK}(\sigma')\right]&=&-i\eta_{[K|[M}S_{N]|L]}\delta(\sigma-\sigma')\\
	\left[S_{MN}(\sigma),P_{L}(\sigma')\right]&=&-iP_{[M}\eta_{N]L}\delta(\sigma-\sigma')\\
	\left[P_{M}(\sigma),P_{N}(\sigma')\right]&=&i(\Sigma_{MN}-f_{MN}{}^LP_L)\delta(\sigma-\sigma')+ i\eta_{MN}\partial_\sigma\delta(\sigma-\sigma')\\
	\left[S_{MN}(\sigma),\Sigma^{LK}(\sigma')\right]&=&-i\delta^{[K}_{M}\Sigma_{N}^{~~L]}\delta(\sigma-\sigma')-i\delta^{[K}_M\delta_N^{L]}\partial_\sigma\delta(\sigma-\sigma') 
\end{array}
\right.}~~ \label{ndPMNnA}
\eea
and the others are zero.
In the left and the right components the current algebras  are written as
\bea
&&{\rm Left}\nn\\
&&{\renewcommand{\arraystretch}{1.3}
\left\{\begin{array}{ccl}
\left[S_{mn}(\sigma),S_{lk}(\sigma')\right]&=&-i\eta_{[k|[m}S_{n]|l]}\delta(\sigma-\sigma')\\
\left[S_{mn}(\sigma),P_{l}(\sigma')\right]&=&-iP_{[m}\eta_{n]l}\delta(\sigma-\sigma')\\
\left[P_{m}(\sigma),P_{n}(\sigma')\right]&=&i\Sigma_{mn}\delta(\sigma-\sigma')+ i\eta_{mn}\partial_\sigma\delta(\sigma-\sigma')\\
\left[P_{\mu}(\sigma),P_{\nu}(\sigma')\right]&=&-if_{\mu\nu}{}^{\lambda}P_\lambda\delta(\sigma-\sigma')+ i\eta_{\mu\nu}\partial_\sigma\delta(\sigma-\sigma')\\
\left[S_{mn}(\sigma),\Sigma^{lk}(\sigma')\right]&=&-i\delta^{[k}_{m}\Sigma_{n}^{~l]}\delta(\sigma-\sigma')-i\delta^{[k}_m\delta_n^{l]}\partial_\sigma\delta(\sigma-\sigma') 
\end{array}
\right.}\label{ndPl}
\\\nn\\
&&{\rm Right}\nn\\
&&{\renewcommand{\arraystretch}{1.3}
\left\{\begin{array}{ccl}
	\left[S_{m'n'}(\sigma),S_{l'k'}(\sigma')\right]&=&
	i\eta_{[k'|[m'}S_{n']|l']}\delta(\sigma-\sigma')\\
	\left[S_{m'n'}(\sigma),P_{l'}(\sigma')\right]&=&iP_{[m'}\eta_{n']l'}\delta(\sigma-\sigma')\\
	\left[P_{m'}(\sigma),P_{n'}(\sigma')\right]&=&-i\Sigma_{m'n'}\delta(\sigma-\sigma')
+ i\eta_{m'n'}\partial_\sigma\delta(\sigma-\sigma')\\
\left[S_{m'n'}(\sigma),\Sigma^{l'k'}(\sigma')\right]&=&i\delta^{[k'}_{m'}\Sigma_{n'}^{~l']}\delta(\sigma-\sigma')
+i\delta^{[k'}_{m'}\delta_{n'}^{l']}\partial_\sigma\delta(\sigma-\sigma') 
\end{array}
\right.}\label{ndPr}
\eea

The background fields $g_{mn}$ and $B_{mn}$ are coupled  to 
the momentum index of the
momentum/winding mode basis.
In order to distinguish  the left/right coordinate
we utilize the notation  $(P_{\underline{\rm m}}, ~
\tilde{P}^{\underline{\rm m}})$ for  
the momentum and the winding mode.
They  are related to the left/right coordinates as
\bea
P_{\underline{M}}=
\left\{
\begin{array}{ccl}
P_{\underline{m}}&=&\displaystyle\frac{1}{\sqrt{2}}(P_m+P_{m'})\\
P_\mu&&\\
\tilde{P}^{\underline{m}}&=&\displaystyle\frac{1}{\sqrt{2}}(P_m-P_{m'})
\end{array}
\right.~~~.\label{puLR}
\eea
The O($D+n$,$D$) invariant metric in this basis is given as
\bea
&&\begin{array}{l}
	~~~~~~~~~~~~~~~~~~~~~~~~~~~~_{\underline{n}}~~~~~~_{\nu}~~~~~~~^{\underline{n}}\\
	\eta_{PP}=	\eta_{\underline{MN}}=\begin{array}{c}_{\underline{ m}}\\_\mu\\^{\underline{ m}}\end{array}
	\left(\begin{array}{ccc}
	&&\delta_{\underline{m}}^{\underline{n}}\\
		&\eta_{\mu\nu}&\\
	\delta_{\underline{n}}^{\underline{m}}	&&\end{array}
	\right)=
	\left(\begin{array}{ccc}
	&&	{\bf 1}\\
		&{\bf 1}&\\
		{\bf 1}	&&\end{array}
	\right)~~~.
\end{array}
\eea
\par
\vskip 6mm 
\section{Geometry of heterotic mega-space} \label{sec:geometry}
\subsection{Background fields}

The gravitational background is introduced which satisfies the orthogonal condition as
\bea
\dd_{\cal A}=E_{\cal A}{}^{\cal M}\dd_{\cal M} ~~,~~
E_{\cal A}{}^{\cal M} E_{\cal B}{}^{\cal N} \eta_{\cal MN}=\eta_{\cal AB}~~~,
\eea
while the momentum part itself also satisfies the orthogonal conditions as
\bea
E_{{A}}{}^{{M}} E_{{B}}{}^{{N}} \eta_{{MN}}=\eta_{{AB}}~~,~~
E_{\underline{A}}{}^{\underline{M}} E_{\underline{B}}{}^{\underline{N}} \eta_{\underline{MN}}=\eta_{\underline{AB}}~~~.
\eea

It is convenient to choose the following gauge for the background field $E_{\cal A}{}^{\cal M}$ including the Lorentz connection $\omega_{\underline{A}}{}^{\underline{MN}}$ \cite{Polacek:2013nla} 
as well as the gravitational vielbein $E_{\underline{A}}{}^{\underline{M}}$ and the non-Abelian gauge field $A_{\underline{m}}{}^\mu$ \cite{Hatsuda:2022zpi}, 
\bea
&&\begin{array}{l}
	~~~~~~~~~~~~~~~~~~~~~~~~~~~~~~~~~~^{MN}~~~~~~~~~~~~~~~~~~~~~~~^{\underline{M}}~~~~~~~~~~_{MN}\\
E_{\cal A}{}^{\cal M}=\begin{array}{c} _{AB}\\ _{\underline{A}}\\ ^{AB}\end{array}
	\left(\begin{array}{ccc}
\delta_{A}^{[M}\delta_{B}^{N]}&0&0\\
\omega_{\underline{A}}{}^{MN}	&E_{\underline{A}}{}^{\underline{M}}&0\\
-\frac{1}{2}\omega^{CAB}\omega_{C}{}^{MN}+r^{ABMN}	&-\omega^{\underline{M}AB}&
\delta^{A}_{[M}\delta^{B}_{N]}
\end{array}
	\right) \label{vielbeinSpSg}
\end{array}
\\
&&\begin{array}{l}
~~~~~~~~~~~~~~~~~~~~~~~~~~~~~~~~~~_{\underline{m}}~~~~~~~~~~~~~~~~~~~~^{\mu}~~~~~~~~^{\underline{m}}\\
E_{\underline{A}}{}^{\underline{M}}=\begin{array}{c}^{\underline{a}}\\_\alpha\\_{\underline{a}}\end{array}
	\left(\begin{array}{ccc}
e_{\underline{m}}{}^{\underline{a}}	&0&0 \\
-e_\alpha{}^\nu A_{\underline{m}\nu}		& e_\alpha{}^\mu &0\\
e_{\underline{a}}{}^{\underline{n}}
(B_{\underline{nm}} -\displaystyle\frac{1}{2}A_{\underline{n} }{}^\nu A_{\underline{m} \nu})	&
e_{\underline{a}}{}^{\underline{n}} A_{\underline{n}}{}^\mu & e_{\underline{a}}{}^{\underline{m}}  \end{array}
	\right)~~~.
\end{array}
\eea
Their inverses are given as
\bea
&&\begin{array}{l}
	~~~~~~~~~~~~~~~~~~~~~~~~~~~~~~~~~~^{AB}~~~~~~~~~~~~~~~~~~~~~~~^{\underline{A}}~~~~~~~~~~_{AB}\\
E^{-1}{}_{\cal M}{}^{\cal A}=\begin{array}{c} _{MN}\\ _{\underline{M}}\\ ^{MN}\end{array}
	\left(\begin{array}{ccc}
\delta^{A}_{[M}\delta^{B}_{N]}&0&0\\
-\omega_{\underline{M}}{}^{AB}	&E_{\underline{M}}{}^{\underline{A}}&0\\
-\frac{1}{2}\omega^{CMN}\omega_{C}{}^{AB}-r^{MNAB}	&\omega^{\underline{A}MN}&
\delta_{A}^{[M}\delta_{B}^{N]}
\end{array}
	\right)
\end{array}\label{Einverse}
\\
&&\begin{array}{l}
~~~~~~~~~~~~~~~~~~~~~~~~~~~~~~~~~~_{\underline{a}}~~~~~~~~~~~~~~~~~~~~~~~^{\alpha}~~~~~~~~^{\underline{a}}\\
E^{-1}{}_{\underline{M}}{}^{\underline{A}}=\begin{array}{c}^{\underline{m}}\\_\mu\\_{\underline{m}}\end{array}
	\left(\begin{array}{ccc}
e_{\underline{a}}{}^{\underline{m}}	&0&0 \\
e_{\underline{a}}{}^{\underline{n}} A_{\underline{n}\mu}		& e_\mu{}^\alpha &0\\
-(B_{\underline{mn}} +\displaystyle\frac{1}{2}A_{\underline{m} }{}^\nu A_{\underline{n} \nu})e_{\underline{a}}{}^{\underline{n}}	&
-A_{\underline{m}}{}^\nu e_{\nu}{}^{\alpha}  & e_{\underline{m}}{}^{\underline{a}}  \end{array}
	\right)~~~.
\end{array}
\eea
The rank four tensor $r^{ABCD}=r^{ABMN}E_{M}{}^{C}E_N{}^D$ is an
anti-symmetric tensor satisfying $r^{ABCD}+r_{CDAB}=0$.
We note that the $\alpha'$ dependence can be restored by the rescaling $A_m{}^\mu ~\to~ \sqrt{\alpha'}A_m{}^\mu$.

\par
\vskip 6mm 
\subsection{Torsion and Bianchi identity}

The current algebra \bref{CAflat} is generalized  in backgrounds as
\bea
\left[\dd_{\cal A}(\sigma),\dd_{\cal B}(\sigma')\right]&=&-iT_{\cal {AB}}{}^{\cal C} \dd_{\cal C}\delta(\sigma-\sigma')
+ i\eta_{\cal {AB}}\partial_\sigma\delta(\sigma-\sigma')~~~.
\label{CABG}
\eea
Torsion $T_{\cal {AB}}{}^{\cal C}$ appears in the current algebra
\bea
T_{\cal {AB}}{}^{\cal C}&=&E_{[{\cal A}|}{}^{\cal M}(D_{\cal M} E_{|{\cal B}]}{}^{\cal N}) E^{-1}{}_{\cal N}{}^{\cal C}
+\frac{1}{2} \eta^{\cal{CD} }E_{\cal D}{}^{\cal M}(D_{\cal M} E_{[{\cal A}}{}^{\cal N})E_{{\cal B}]}{}^{\cal L}\eta_{\cal NL}
\label{Tabupc}\\
&&+E_{{\cal A}}{}^{\cal M} E_{{\cal B}}{}^{\cal N} f_{\cal MN}{}^{\cal L} E^{-1}{}_{{\cal L}}{}^{\cal C}~~~.\nn
\eea
Lowering the last index of the torsion gives the totally anti-symmetric rank three tensor
\bea
	T_{\cal {ABC}}&\equiv&T_{\cal {AB}}{}^{\cal D}\eta_{\cal DC}~=~\frac{1}{3!}T_{\cal {[ABC]}}~=~
	\frac{1}{2}E_{[{\cal A}|}{}^{\cal M}(D_{\cal M} E_{|{\cal B}}{}^{\cal N}) E_{{\cal C}]}{}_{\cal N}
	+E_{{\cal A}}{}^{\cal M} E_{{\cal B}}{}^{\cal N}E_{{\cal C}}{}^{\cal L} f_{\cal MNL}~~~.\nn\\
	\label{TABC}
\eea
The second term in the right hand side of \bref{Tabupc} is the stringy contribution caused by
the Schwinger term
in the current algebra \bref{CABG} as
\bea
&&E_{\cal A}{}^{\cal M}(\sigma)E_{\cal B}{}^{\cal N}(\sigma')\eta_{\cal MN}~\partial_\sigma \delta(\sigma-\sigma')\nn\\
&&~~~=
(1-K)\big(
\eta_{\cal AB}~\partial_\sigma \delta(\sigma-\sigma')+E_{\cal A}{}^{\cal M}\partial_\sigma E_{\cal B}{}^{\cal N}\eta_{\cal MN}~ \delta(\sigma-\sigma')
\big)\nn\\
&&~~~~~~~+K\big(
\eta_{\cal AB}~\partial_\sigma \delta(\sigma-\sigma')-(\partial_\sigma E_{\cal A}{}^{\cal M}) E_{\cal B}{}^{\cal N}\eta_{\cal MN} ~\delta(\sigma-\sigma')
\big)~~~
\eea
with an arbitrary constant $K$. For the antisymmetric property with respect to $_{\cal AB}$ we choose  $K=1/2$.
It is mentioned that the hyperfunction includes an ambiguity as
\bea
f(\sigma)g(\sigma')~\partial_\sigma\delta(\sigma-\sigma')&=&
fg(\sigma)~\partial_\sigma\delta(\sigma-\sigma')+f (\partial_\sigma g) ~\delta(\sigma-\sigma')\nn\\
&=&
fg(\sigma')~\partial_\sigma\delta(\sigma-\sigma')-(\partial_\sigma f ) g ~\delta(\sigma-\sigma')~~~,
\eea
then evaluating $\sigma$-derivative by \bref{dsigma} leads to the second term of \bref{Tabupc}.
This term vanishes under the strong section condition as
$\eta^{\cal CD}(D_{\cal D}E)\dd_{\cal C}=
\eta^{\cal MN}(D_{\cal M}E)\dd_{\cal N}
=0$
and the gravitational part of the torsion reduces into the usual gravity theory in \bref{RLr}.


The Bianchi identity is obtained as follows \cite{Hatsuda:2014qqa}
\bea
&&\left[\dd_{\cal A}(\sigma), \left[\dd_{\cal B}(\sigma'),\dd_{\cal C}(\sigma'')\right]\right]
+
\left[\dd_{\cal B}(\sigma'), \left[\dd_{\cal C}(\sigma''),\dd_{\cal A}(\sigma)\right]\right]
+
\left[\dd_{\cal C}(\sigma''), \left[\dd_{\cal A}(\sigma),\dd_{\cal B}(\sigma')\right]\right]\nn\\
&&~~=~-{\cal I}_{\cal ABC}{}^{\cal E}\dd_{\cal E}\delta(\sigma-\sigma')\delta(\sigma-\sigma'')~=~ 0~~~.
\label{Bianchi}
\eea
Again lowering the index of ${\cal I}_{\cal ABC}{}^{\cal D}$ becomes totally anti-symmetric by using the ambiguity caused by the 
$\partial_\sigma \delta(\sigma)$ in the current algebra \bref{Bianchi} as
\bea
{\cal I}_{\cal ABCD}&\equiv&{\cal I}_{\cal ABC}{}^{\cal E}\eta_{\cal ED}~=~\displaystyle\frac{1}{4!}{\cal I}_{[{\cal ABCD}]}
~=~\frac{1}{3!}\left(
D_{[{\cal A}}T_{\cal BCD]} +\frac{3}{4}T_{\cal [AB}{}^{\cal E}T_{\cal CD]E}
\right)~=~0~~~.
\eea
\par
\vskip 6mm
\subsection{Effective gravity theory}

The effective gravity theory is obtained by the zero-mode limit (
particle limit) analogously to \bref{RLr}.
The zero-mode limit of the stringy momentum current is denoted by
$\dd_{\cal A}(\sigma)\to\nabla_{\cal A}$.
The commutator of the two $\nabla_P$ is given as
\bea
\left[\nabla_A,\nabla_B\right]&=&\frac{1}{i}T_{AB}{}^C\nabla_C+ \frac{1}{2i}R_{AB}{}^{CD}S_{CD}~~~.\label{PPR}
\eea
Let us evaluate the zero-mode limit of 
the stringy algebra in the doubled heterotic space 
	\bref{CABG} focusing on the commutator of two $P$'s. 
Because of $\dd_{\cal A}=E_{\cal A}{}^{\cal M}\dd_{\cal M}$, 
the torsion $T_{\cal AB}{}^{\cal C}$ in \bref{CABG}
is written by 
the torsion with mixed indices $t_{\cal AB}{}^{\cal M}$ 
and $(E^{-1})_{\cal M}{}^{\cal A}$ in \bref{Einverse} as
\bea
\left[\dd_{\cal A}(\sigma),\dd_{\cal B}(\sigma')\right]
&=&\frac{1}{i}t_{\cal AB}{}^{\cal M}\dd_{\cal M}\delta(\sigma-\sigma')+i\eta_{\cal AB}\partial_\sigma \delta(\sigma-\sigma')\nn\\
&=&\frac{1}{i}t_{\cal AB}{}^{\cal M}(E^{-1})_{\cal M}{}^{\cal C}\dd_{\cal C}\delta(\sigma-\sigma')+i\eta_{\cal AB}\partial_\sigma \delta(\sigma-\sigma')
~~~.
\eea 
Schematically, the $PP$ component of \bref{CABG}, with
$ \dd_{\mathcal A}=(S_{AB},P_A,\Sigma^{AB}),$
takes the following form.
\bea
&&\left[P(\sigma),P(\sigma')\right]
~=~\frac{1}{i}(T_{PP}{}^{P}P+T_{PP}{}^SS+
T_{PP}{}^\Sigma \Sigma)\delta(\sigma-\sigma')+i\eta_{AB}\partial_\sigma\delta(\sigma-\sigma')~~~\label{PPTP}\\
&&
\left\{
\begin{array}{ccl}
T_{PP}{}^{P}&=&t_{PP}{}^P(E^{-1})_P{}^P+t_{PP}{}^\Sigma(E^{-1})_\Sigma{}^P
~\Rightarrow~0~{\rm torsion~constraint}\\
T_{PP}{}^{S}&=&t_{PP}{}^P(E^{-1})_P{}^S+t_{PP}{}^S{\bf 1}_S{}^S
+t_{PP}{}^\Sigma(E^{-1})_\Sigma{}^S\\
&&~~~~~~~~~~~~~~~~~~~~~~~~~~~~~~~~~~~~
~\Rightarrow~R_{PP}{}^S~{\rm Riemann~tensor}\\
T_{PP}{}^{\Sigma}&=&t_{PP}{}^P(E^{-1})_P{}^\Sigma+t_{PP}{}^S(E^{-1})_S{}^\Sigma
+t_{PP}{}^\Sigma(E^{-1})_\Sigma{}^\Sigma
\end{array}
\right.
\label{Tgravity}
\eea
In component form, the mixed-index torsions are given by
\bea
\left\{
\begin{array}{ccccl}
t_{PP}{}^{P}&=&t_{AB}{}^M&=&E_{[A|}{}^N\partial_N E_{|B]}{}^M
-\frac{1}{2}E_{[A|}{}^L(\partial_NE_{|B]L})\eta^{NM}
-\omega_{[AB]}{}^CE_C{}^M\\&&&& +E_{A}{}^NE_B{}^Lf_{NL}{}^M \\
t_{PP}{}^{S}&=&t_{AB}{}^{MN}&=&E_{[A|}{}^L\partial_L\omega_{|B]}{}^{MN}
+\omega_{[A}{}^{LM}\omega_{B]L}{}^{N}-\omega_{[AB]}{}^C\omega_{C}{}^{MN}
\\
t_{PP}{}^{\Sigma}&=&t_{ABMN}&=&-E_{[A|M}E_{|B]N}~~~.
\end{array}
\right.
\label{TgravityAB}
\eea
 In the low energy effective theory limit (particle limit), both the Schwinger term and the non-degenerate generator $\Sigma^{AB}$ are absent.

At first we solve the torsion constraint
following \cite{Siegel:1999ew,Polacek:2013nla}
\bea
T_{PPP}~=~T_{ABC}~=~T_{AB}{}^D\eta_{DC}~=~\frac{1}{3!}T_{[ABC]}
~=~0~~\label{torsionconstraint}~~~.
\eea
Using  \bref{Tgravity} this constraint can be expressed as
\bea
t_{PP}{}^P~=~-t_{PP}{}^\Sigma(\omega )_\Sigma{}^PE_P{}^P \label{tppp}
\eea
which provides a convenient relation for computing the curvature.
More explicitly, substituting \bref{TABC} and \bref{vielbeinSpSg} into the torsion constraint \bref{torsionconstraint} yields
\bea
\omega_{[\underline{ABC}]}&=&E_{[\underline{A}|}{}^{\underline{N}}(\partial_{\underline{N}} E_{|\underline{B}}{}^{\underline{M}})E_{\underline{C}]\underline{M}}
+2E_{\underline{A}}{}^{\mu}E_{\underline{B}}{}^{\nu}E_{\underline{C}}{}^{\rho}f_{\mu\nu\rho}~~~.\label{omega}
\eea
Since the only independent components are  $\omega_{\underline{A}bc}$ and $ \omega_{\underline{A}b'c'}$,
the following relations hold 
\bea
\omega_{\underline{A}\underline{bc}}&=&\frac{1}{2}(\omega_{\underline{A}{bc}}+\omega_{\underline{A}bc'}+ \omega_{\underline{A}b'c}+ \omega_{\underline{A}b'c'})
~=~\frac{1}{2}(\omega_{\underline{A}{bc}}+ \omega_{\underline{A}b'c'})
\nn\\
\omega_{\underline{A}\underline{b}}{}^{\underline{c}}&=&\frac{1}{2}(\omega_{\underline{A}{bc}}-\omega_{\underline{A}bc'}+ \omega_{\underline{A}b'c}- \omega_{\underline{A}b'c'})
~=~\frac{1}{2}(\omega_{\underline{A}{bc}}- \omega_{\underline{A}b'c'})\nn\\
\omega_{\underline{A}}{}^{\underline{b}}{}_{\underline{c}}&=&\frac{1}{2}(\omega_{\underline{A}{bc}}+\omega_{\underline{A}bc'}- \omega_{\underline{A}b'c}- \omega_{\underline{A}b'c'})
~=~\frac{1}{2}(\omega_{\underline{A}{bc}}- \omega_{\underline{A}b'c'})\nn\\
\omega_{\underline{A}}{}^{\underline{bc}}&=&\frac{1}{2}(\omega_{\underline{A}{bc}}-\omega_{\underline{A}bc'}- \omega_{\underline{A}b'c}+ \omega_{\underline{A}b'c'})
~=~\frac{1}{2}(\omega_{\underline{A}{bc}}+ \omega_{\underline{A}b'c'})\nn\\
\Rightarrow &&\omega_{\underline{A}\underline{bc}}= \omega_{\underline{A}}{}^{\underline{bc}}~~,~~
\omega_{\underline{A}\underline{b}}{}^{\underline{c}}= \omega_{\underline{A}}{}^{\underline{b}}{}_{\underline{c}}~~~.
\label{windices}
\eea
The number of degrees of freedom of $\omega_{[\underline{ABC} ] }$ is as follows:
For the $\underline{A}=(_{\underline{a}}, ~^{\underline{a}})$ part, the number for totally antisymmetric indices using \bref{windices} is $\frac{1}{2\cdot 3!}[2D(2D-1)(2D-2)]
=\frac{1}{3}D(2D-1)(D-1)$.
On the other hand, the number for $\omega_{\underline{abc}}=\omega^{\underline{abc}} $ is $\frac{1}{3!}D(D-1)(D-2)$ 
and that for $\omega_{\underline{ab}}{}^{\underline{c}}=\omega^{\underline{ab}}{}_{\underline{c}} $ is $\frac{1}{2}D^2(D-1)$, and summing up both numbers gives
$\frac{1}{3!}D(D-1)(D-2)+\frac{1}{2}D^2(D-1)=\frac{1}{3}D(2D-1)(D-1)$, which coincides with the number for the $\underline{A}=(_{\underline{a}}, ~^{\underline{a}})$ part of $\omega_{[\underline{ABC} ] }$.  

In this paper, to compare our formulation with the conventional
heterotic supergravity action, we assume that the heterotic supergravity
fields depend only on the ordinary spacetime coordinates and not on the
additional coordinates of mega-space,
$\partial_{\underline{M}}=(\partial_{\underline{m}}, \partial_\mu=0,
\partial^{\underline{m}}=0)$. 
This is the solution of the section condition.
Among three kinds of momenta $P_{\underline{A}}=(P_{\underline{a}}, P_{\alpha}, P^{\underline{a}})$,
	only commutator of two $P_{\underline{a}}$'s gives the heterotic extension of \bref{PPR}.
Three torsion constraints $T_{\underline{ab}}{}^{\underline{c}}=T_{\underline{ab}}{}^\gamma=T_{\underline{abc}}=0$ as coefficients of $P_{\underline{c}}, P_\gamma, P^{\underline{c}}$ in \bref{PPTP} determine $\omega$'s.
As a result, the three types of Lorentz connections in \bref{omega} correspond to the three field strengths ${H}_{\underline{abc}}$, $c_{\underline{ab}}{}^{\underline{c}}$, and $F_{\underline{ab}}{}^\gamma$ respectively, and are given by
\bea
&
\frac{1}{2}\omega_{[\underline{ABC}]}~=~
{\renewcommand{\arraystretch}{1.3}
\left\{\begin{array}{ccl}
\omega_{\underline{abc}}+\omega_{\underline{cab}}+\omega_{\underline{bca}}
&=& {H}_{\underline{abc}}~=~
  e_{\underline{a}}{}^{\underline{m}} e_{\underline{b}} {}^{\underline{n}} e_{\underline{c}} {}^{\underline{l}}{H}_{\underline{mnl}}~\\
\omega_{\underline{ab}}{}^{\underline{c}}+\omega^{\underline{c}}{}_{\underline{ab}}
+\omega_{\underline{b}}{}^{\underline{c}}{}_{\underline{a}}
&=&\frac{1}{4}\eta^{\underline{cd}}c_{[\underline{abd}]}
\\
\omega^{\gamma}{}_{\underline{ab}}&=& F_{\underline{ab}}{}^\gamma~=~ e_{\underline{a}}{}^{\underline{m}} e_{\underline{b}}{}^{\underline{n}} 
F_{\underline{mn}}{}^{\mu}e_\mu{}^\gamma
\end{array}\right.
}\label{torsions}&
\\
\nn\\
&\left\{ \begin{array}{ccl}
{H}_{\underline{mnl}}&=&\frac{1}{2}\left(
\partial_{[\underline{m}}B_{\underline{nl}]}+A_{[\underline{m}}{}^{\mu}\partial_{\underline{n}}A_{\underline{l}]\mu}
\right)+A_{\underline{m}}{}^{\mu} A_{\underline{n}}{}^{\nu} A_{\underline{l}}{}^{\lambda} f_{\mu\nu\lambda}  \\
c_{\underline{ab}}{}^{\underline{c}}&=&
e_{[\underline{a}|}{}^{\underline{m}} (\partial_{\underline{m}} e_{|\underline{b}]}{}^{\underline{n}} )e_{\underline{n}}{}^{\underline{c}}\\
F_{\underline{mn}}{}^{\mu}&=&\partial_{[\underline{m}}A_{\underline{n}]}{}^{\mu}
+A_{\underline{m}}{}^\nu A_{\underline{n}}{}^\rho f_{\nu\rho}{}^\mu  \end{array}\right.&
\label{fieldstrength}~~~.
\eea
The symmetric part of 
the Lorentz connection $\omega_{\underline{ab}}{}^{\underline{c}}$
is not determined by the torsion constraint which is totally
anti-symmetric.
This structure is reminiscent of the curvature tensor in DFT \cite{Jeon:2011cn, Hohm:2011si}.
We assume the symmetric part as the same as the non-doubled geometry in the \bref{Aomega} as
\bea
\omega_{\underline{ab}}{}^{\underline{c}}&=&\frac{1}{2}
(c_{\underline{ab}}{}^{\underline{c}} +c^{\underline{c}}{}_{(\underline{ab})})~~~.\label{omegac}
\eea

Next we calculate the Riemann curvature tensor $R_{PP}{}^S$ in \bref{Tgravity}.
Substituting \bref{tppp} and \bref{Einverse}  
into the second line of \bref{Tgravity}
yields
\bea
R_{PP}{}^S~=~T_{PP}{}^{S}&=& 
t_{PP}{}^{S} {\bf 1}_{S}{}^S
+t_{PP}{}^{\Sigma}\left(-\omega _\Sigma{}^P E_P{}^P (E^{-1})_P{}^S 
+E^{-1}{}_{\Sigma}{}^S\right)
~~~.\label{RPPS}
\eea
It is noted that $t_{PP}{}^S$ corresponds to the conventional Riemannian curvature tensor in the non-doubled space \bref{RLr}.
The rank four tensor $r_{ABCD}$ enters through 
 the last term of the right hand side of \bref{RPPS}, $E^{-1}{}_\Sigma{}^S$
 in \bref{Einverse}.
$r_{ABCD}$ is antisymmetric under the exchange of the index pairs $_{AB}$ and $_{CD}$,
which renders $R_{ABCD}$ symmetric under the same exchange.
In component form, the curvature tensor \bref{RPPS} is given by
\bea
R_{{AB}}{}^{CD}&=&T_{{AB}}{}^{CD}\nn\\
&=&E_{[{A}|}{}^{{L}}\partial_{{L}}\omega_{|{B}]}{}^{{CD}}
+\omega_{[{A}}{}^{EC}\omega_{{B}]E}{}^{D}
-\omega_{[{AB}]}{}^{E}\omega_E{}^{CD}
-\frac{1}{2}\omega_{EAB}\omega^{ECD}
+r_{AB}{}^{CD}~~~.\nn\\
\label{TPPS}
\eea

The Bianchi identity of three $P_A$'s gives
\bea
[P_{[A},[P_B,P_{C]}]]&=&[P_{[A},\frac{1}{2i}R_{BC]}{}^{DE}S_{DE}]]\nn\\
&=&\frac{1}{2i}
\left(
[P_{[A},R_{BC]}{}^{DE}]S_{DE}-R_{[BCA]}{}^{D}P_{D}\right)~=~0~~~\nn\\
&\Rightarrow
& D_{[A}R_{BC]}{}^{DE}~=~0~,~R_{[ABC]}{}^{D}=0~~~.\label{RABC0}
\eea
The second identity and the anti-symmetric property of the two indices lead to the following property
\bea
&&R_{[ABC]}{}^{D}=0~,~R_{ABCD}=-R_{BACD}~,~R_{ABCD}=-R_{ABDC}\nn\\
&&\to
\left\{
\begin{array}{lcc}
R_{[ABC]D}=0=R_{[AB]CD}+R_{[BC]AD}+R_{[CA]BD}&\cdots&(1)\\
R_{[BCD]A}=0=R_{[BC]DA}+R_{[CD]BA}+R_{[DB]CA}&\cdots&(2)\\
R_{[CDA]B}=0=R_{[CD]AB}+R_{[DA]CB}+R_{[AC]DB}&\cdots&(3)\\
R_{[DAB]C}=0=R_{[DA]BC}+R_{[AB]DC}+R_{[BD]AC}&\cdots&(4)\\
\end{array}	\right.
\nn\\
&&(1)+(2)-(3)-(4)=0~\Rightarrow~ R_{ABCD}=R_{CDAB}~~~.
\eea
The physical Riemannian tensor is 
\bea
\raisebox{0.7ex}
{\ydiagram{2,2}}&:&\frac{1}{2}
(R_{[AB][CD]}+R_{[CD][AB]})\neq 0\label{Riemanntensor}
\eea
and trace parts 
$\raisebox{0.0ex}{\ydiagram{2}}$ and $\bullet$ are obtained by  contractions as $R_{AC}=R_{ABCD}\eta^{BD}$ and 
$R=R_{AC}\eta^{AC}$.

The Riemannian curvature tensor in \bref{Riemanntensor} is 
\bea
R_{ABCD}
	&=&\frac{1}{2}
\Big(
E_{[A|}{}^{L}(\partial_{L}\omega_{|B]CD})
+E_{[C|}{}^{L}(\partial_{L}\omega_{|D]AB})
+\omega_{[A|C}{}^{E}\omega_{|B]DE}
+\omega_{[C|A}{}^{E}\omega_{|D]BE}\Big. \nn\\
&&~~~\Big.
-\omega_{[AB]}{}^{E}\omega_{ECD}
-\omega_{[CD]}{}^{E}\omega_{EAB}
-\omega^{E}{}_{AB}\omega_{ECD}\Big)~~~.\label{RABCD}
\eea
As in \bref{windices}, 
distinguishing the $V^A$ index as either a lower index 
$V_{\underline{a}}$ 
	or an upper index $V^{\underline{a}}$ 
	determines the corresponding index structure of the field strength in \bref{torsions}. 
	Expanding the field strength in terms of these components, one finds that $R_{\underline{abc}}{}^{\underline{d}}$ corresponds to the  Riemannian tensor for the heterotic supergravity
\bea
R_{\underline{abc}}{}^{\underline{d}}&=& 
R_{0:\underline{abc}}{}^{\underline{d}}+\Delta R_{0:\underline{abc}}{}^{\underline{d}}
+\displaystyle\frac{1}{3^2}
\left(
H_{\underline{ce}[\underline{a}|}H_{|\underline{b}]}{}^{\underline{de}}
-H_{[\underline{ab}]\underline{e}}H_{\underline{c}}{}^{\underline{de}}
-\frac{1}{2}H_{\underline{ab}\underline{e}}H_{\underline{c}}{}^{\underline{de}}
 \right)
-\frac{1}{2}F_{\underline{ab}}{}^{\gamma}
F_{\underline{c}}{}^{\underline{d}}{}_{\gamma}\nn\\
R_{0:\underline{abc}}{}^{\underline{d}}&=& 
e_{[\underline{a}|}{}^{\underline{m}}\partial_{\underline{m}}\omega_{|\underline{b}]\underline{c}}{}^{\underline{d}}
+
\omega_{[\underline{a}|\underline{c}}{}^{\underline{e}}
\omega_{|\underline{b}]}{}^{\underline{d}}{}_{\underline{e}}
-\omega_{[\underline{ab}]}{}^{\underline{e}}
\omega_{\underline{ec}}{}^{\underline{d}}
\nn\\
\Delta R_{0:\underline{abc}}{}^{\underline{d}}&=&-\frac{1}{2}\omega^{\underline{e}}{}_{\underline{ab}}
\omega_{\underline{ec}}{}^{\underline{d}}
+r_{\underline{abc}}{}^{\underline{d}}
\eea
where  $R_{0:\underline{abc}}{}^{\underline{d}}$ is the conventional form \bref{RLr} and $\Delta R_{0:\underline{abc}}{}^{\underline{d}}$  is the stringy contribution originated from $t_{PP}{}^\Sigma$  in \bref{RPPS}.
Lowering the index $\underline{d}$ as $R_{\underline{abcd}}=R_{\underline{abc}}{}^{\underline{e}}\eta_{\underline{ed}}$ and then symmetrizing the indices $\underline{ab}$ and $\underline{cd}$, we obtain  the Riemannian curvature tensor in the doubled heterotic space as
\bea
R_{\underline{abcd}}&=&\frac{1}{2}(
R_{\underline{abcd}}+ R_{\underline{cdab}})\nn\\
&=&\frac{1}{2}\left(
R_{0:\underline{abcd}}+ R_{0:\underline{cdab}}+
\Delta R_{0:\underline{abcd}}+ \Delta R_{0:\underline{cdab}}
\right)\nn\\
&&+\displaystyle\frac{1}{2\cdot 3^2}\left(
2H_{\underline{ce}[\underline{a}|}H_{|\underline{b}]\underline{d}}{}^{\underline{e}}
-4H_{\underline{e}\underline{ab}}H_{\underline{cd}}{}^{\underline{e}}
-H_{\underline{abe}}H_{\underline{cd}}{}^{\underline{e}} \right)\nn\\
&&-\frac{1}{2}F_{\underline{ab}}{}^{\gamma}F_{\underline{cd}}{}_{\gamma}~~~
\eea
\bea
\frac{1}{2}(R_{0:\underline{abcd}}+ R_{0:
	\underline{cdab}})
&=&\displaystyle\frac{1}{2}\left(
D_{[\underline{a}}\omega_{\underline{b}]\underline{c}}{}_{\underline{d}}
+D_{[\underline{c}}\omega_{\underline{d}]}{}_{\underline{ab}}\right)
\nn\\
\frac{1}{2}(\Delta R_{0:\underline{abcd}}+ \Delta R_{0:\underline{cdab}})
&=&
-\frac{1}{2}\omega^{\underline{e}}{}_{\underline{ab}}
\omega_{\underline{e}}{}_{\underline{cd}}	~~~.\nn
\eea
The additional $\omega^2$ term in $\Delta R_{0:\underline{abcd}}$ 
is required to make it T-duality covariant 
and to ensure consistency with the $H^2$ and $F^2$ terms.

The Ricci tensor in the doubled heterotic space is given by
\bea
R_{\underline{ac}}&=&
R_{\underline{abc}}{}^{\underline{b}}
~=~R_{0:\underline{ac}}
+\Delta R_{0:\underline{ac}}-\displaystyle\frac{1}{3!}
H_{\underline{abd}}H_{\underline{c}}{}^{\underline{bd}}
-\frac{1}{2}F_{\underline{ab}}{}_{\gamma}F_{\underline{c}}{}^{\underline{b}\gamma}~~~
\eea
\bea
R_{0:\underline{ac}}
&=&\displaystyle\frac{1}{2}
D_{(\underline{a}|}\left\{\frac{1}{\bf e} \partial_{\underline{m}}
({\bf e} e_{|\underline{c})}{}^{\underline{m}})\right\}
-\frac{1}{2} D_{\underline{b}} c^{\underline{b}}{}_{(\underline{ac})}
\nn\\
\Delta R_{0:\underline{ac}}
&=&
-\frac{1}{2}\omega^{\underline{eb}}{}_{\underline{a}}
\omega_{\underline{ebc}}	~~~.\nn
\eea

	The scalar curvature is given by
\bea
	R&=&R_{\underline{ab}}{}^{\underline{ab}}
	~=~R_0+\Delta R_{0}-\frac{1}{3!}H_{\underline{abe}}H^{\underline{abe}}
	-\frac{1}{2}F_{\underline{ab}}{}_{\gamma}F^{\underline{ab}}{}^{\gamma}\\
R_0&=&
2	D_{\underline{a}}\left\{\frac{1}{\bf e} \partial_{\underline{m}}
	({\bf e} e^{\underline{am}})\right\}
	\nn\\
	\Delta R_{0}
	&=&
	-\frac{1}{2}\omega^{\underline{abc}}
	\omega_{\underline{abc}}	~~~.\nn
\eea
{After imposing the physical section condition and taking the particle limit,
we supplement the curvature contribution with the standard spacetime dilaton
terms.
This is based on Siegel's treatment \cite{Siegel:1993xq,
Siegel:1993th, Siegel:1993bj} in which a redefined dilaton density is
introduced to compensate for the nontrivial transformation of the spacetime
integration measure under T-duality.
This gives the following heterotic supergravity Lagrangian
in $D$ spacetime dimensions%
\footnote{
In the doubled formalism of DFT, the generalized dilaton $d$, defined by
$e^{-2d}=\sqrt{-g}\,e^{-2\phi}$, is commonly incorporated into the connection
through the compatibility condition $\nabla_A d=0$, so that the dilaton
contribution is encoded in the generalized scalar curvature
\cite{Jeon:2011cn,Hohm:2011si}.
In the present construction, the dilaton contribution is introduced after
imposing the physical section condition, at the level of the $D$-dimensional spacetime
action.
We thank Yuho Sakatani for bringing this point to our attention.
}:
\begin{align}
\mathcal{L} =& \  
\sqrt{-g} \, e^{-2\phi} 
\left[
\frac{1}{2} 
R 
+ 4 (\del \phi)^2
\right]
\notag \\
=& \ 
\sqrt{-g} \, e^{-2\phi}
\left[
\frac{1}{2}(R_0+\Delta R_{0})
+ 4 (\del \phi)^2
-\frac{1}{12} H_{\underline{abc}} H^{\underline{abc}}
-\frac{\alpha'}{4}F_{\underline{ab}}{}_{\gamma}
F^{\underline{ab}}{}^{\gamma}
\right].
\end{align}
Here we have recovered $\alpha'$ explicitly.
}

We will show that the rank four totally antisymmetric part, 
which must be zero, 
corresponds to the  Green-Schwarz condition
\bea
\raisebox{2.5ex}
{\ydiagram{1,1,1,1}}~~&:&R_{[ABCD]}=0  \label{GSc}
\eea
The totally anti-symmetric part of the curvature tensor in \bref{GSc} is given by
\bea
R_{[ABCD]}&=&{2}E_{[A|}{}^M\partial_M\omega_{|BCD]}
-{2} \omega_{[AB|E}\omega_{|CD]}{}^{E}
-{2} \omega_{[AB|E}\omega^E{}_{|CD]}
-\frac{1}{2}\omega^E{}_{[AB|}\omega_{E|CD]}\nn\\
&=&0~~~.
\eea
The four momentum-mode-indices part of the above condition becomes
\bea
R_{[\underline{abcd}]}&=&
\frac{2}{3}e_{[\underline{a}|}{}^{\underline{m}}
\partial_{\underline{m}}{H}_{|\underline{bcd}]}
-\frac{1}{2}c_{[\underline{ab}|}{}^{\underline{e}}{H}_{|\underline{cd}]\underline{e}}
-c^{\underline{e}}{}_{[\underline{ab}|}H_{|\underline{cd}]\underline{e}}
-\frac{\alpha'}{2}F_{[\underline{ab}|}{}^\gamma
F_{|\underline{cd}]}{}_\gamma\nn\\
&=&\frac{2}{3}
D_{[\underline{a}}{H}_{\underline{bcd}]}
-c^{\underline{e}}{}_{[\underline{ab}}H_{\underline{cd}]\underline{e}}
-\frac{\alpha'}{2}F_{[\underline{ab}|}{}^\gamma
F_{|\underline{cd}]}{}_\gamma~=~0
~~~.\label{GScond}
\eea
By redefining a covariant derivative 
\bea
{\cal D}_{[\underline{a}}{H}_{\underline{bcd}]}&=&
D_{[\underline{a}}{H}_{\underline{bcd}]}-
\frac{3}{2}c^{\underline{e}}{}_{[\underline{ab}}{H}_{\underline{cd}]\underline{e}}\nn\\&=&
e_{[\underline{a}}{}^{\underline{m}}\partial_{\underline{m}}
H_{\underline{bcd}]}
-\frac{3}{2}c_{[\underline{ab}}{}^{\underline{e}}H_{\underline{cd}]\underline{e}}-\frac{3}{2}H_{[\underline{ab}|\underline{e}}c^{\underline{e}}{}_{|\underline{cd}]},
\eea
this is a Green-Schwarz-like condition
\bea
{\cal D}_{[\underline{a}}{H}_{\underline{bcd}]}&=&\frac{3\alpha'}{4}
F_{[\underline{ab}|}{}^\gamma F_{|\underline{cd}]}{}_\gamma~~~.
\eea
The coefficient of $F^2$-term depends on the
 normalization of the gauge group $P_\mu$.


\section{Conclusion and discussions} \label{sec:conclusion}

In this paper, we investigated the structure of Lorentz and gauge
connections in heterotic supergravities in a T-duality covariant manner.
We utilized a generalization of the Pol\'a\v{c}ek-Siegel formalism and
described the relevant geometry in the mega-space.  
We incorporated the heterotic non-Abelian gauge sector together with the local Lorentz
sector into the affine current algebra.
As in the original Pol\'a\v{c}ek-Siegel construction, the additional generator $\Sigma$ is
required for the closure of the current algebra and for the consistency
of the Jacobi identities. 
In this enlarged algebraic framework, the
doubled vielbein, the Lorentz connection, and the gauge connection are
organized into the geometric data in the mega-space.

We analyzed the extended current algebra in the mega-space in detail.
By imposing the torsionless condition and the physical section
condition, we extracted the physical spacetime components of the
connections. 
In particular, we explicitly derived the components of the
spacetime spin connection and the Yang-Mills field strength from the
torsion constraints. 
The resulting expressions reproduce the expected
structures of heterotic supergravities. 
Namely, the Lorentz and gauge connections appear in a parallel way, while their different geometric
origins are still kept manifest. 
This provides a current algebraic explanation of why the Yang-Mills and Lorentz Chern--Simons terms enter
the Green-Schwarz modification of the $H$-flux in heterotic supergravity. 

We also derived the Bianchi identity from the Jacobi identity of the
stringy covariant derivatives in the mega-space. 
This result shows that the Green-Schwarz anomaly cancellation condition
is encoded in the consistency of the current algebra. 
In this sense, the Green-Schwarz mechanism has a natural geometric
interpretation in the mega-space formulation. 
The Lorentz Chern--Simons structure is not introduced as
an external ingredient, but is obtained from the Lorentz part of the
affine geometry.  
At the same time, the non-Abelian gauge Chern--Simons
structure is incorporated through the gauge current algebra. 
The combination of these two sectors gives the standard heterotic structure
in a unified but geometrically distinguishable way.

We further studied the gauge structure in the mega-space.
This gauge structure is similar to the one appearing in gauged DFT
\cite{Hohm:2011ex, Grana:2012rr, Hatsuda:2022zpi, Mori:2024sfg}, but it is
extended so that the Lorentz and gauge connections are treated within
the same affine current algebraic framework. 
We also showed that the Lorentz block of the mega-vielbein transforms as the ordinary Lorentz
connection.
This supports the interpretation that the local Lorentz
sector is incorporated geometrically from the beginning, rather than
being added only as a part of an internal gauge sector. 
In this respect, our approach is complementary to the gauged DFT formulation in
which the local Lorentz symmetry is treated as part of the gauge sector
\cite{Bedoya:2014pma}. 

Our approach also provides a complementary perspective to recent
mega-space formulations of the tensor hierarchy and generalized
Green-Schwarz transformations \cite{Hassler:2023axp,Hassler:2024yis,Gitsis:2025clo}.
While those works clarify the algebraic structure of higher-form
symmetries and all-order Green-Schwarz transformations, the present
construction emphasizes the affine current algebra underlying the
Lorentz and gauge sectors. 
It would be useful to understand more
precisely how the curvature tensors obtained in the present mega-space
formalism are related to the hierarchy of curvatures and to all-order
duality covariant Green-Schwarz transformations. 

It would also be interesting to clarify the precise relation between
the present current algebraic construction and the generalized
Bergshoeff--de Roo identification used in $\alpha'$-corrected DFT
\cite{Baron:2018lve, Baron:2020xel, Gitsis:2025clo, Gitsis:2026ufz}.
Our results suggest that the parallel treatment of the torsionful
Lorentz connection and the Yang-Mills connection may have a natural
origin in the affine geometry of the mega-space. 

The Pol\'a\v{c}ek-Siegel formalism is a useful scheme for exploring
$\alpha'$-corrections to supergravities from a T-duality covariant
viewpoint.  
There are some directions for future work. 
In the present current-algebraic approach to heterotic mega-space, we have not been able to derive the $\alpha'\operatorname{tr}(R\wedge R)$ term associated with the Green–Schwarz anomaly cancellation mechanism. 
It is therefore important to clarify how this term can be reproduced
within the current algebra framework.
It would be interesting to apply the same method to more general extended geometries and to clarify whether similar current algebraic structures underlie other duality covariant formulations. 
We leave these issues for future work.

\subsection*{Acknowledgments}
{The authors would like to thank Yuho Sakatani for useful comments on
mega-space framework of heterotic theories.
}
This work is supported in part by Grant-in-Aid for Scientific Research (C), JSPS KAKENHI
Grant Numbers JP22K03603 (M.~H and M.~Y), JP20K03604 (M.~H) and
JP25K07324 (M.~H, S.~S and M.~Y).

\end{document}